\documentclass[aps,prb,twocolumn]{revtex4}

\usepackage{amsmath}
\usepackage{amsfonts}
\usepackage{amssymb}
\usepackage{txfonts}
\usepackage{bm}
\usepackage{tabularx}
\usepackage{graphicx,color}
\usepackage{multirow}

\def\Vec#1{\bm{#1}}
\def\Hc2{H_\mathrm{c2}}
\def\Bc2{B_\mathrm{c2}}
\def\Tc{T_\mathrm{c}}

\begin{document}

\title{Modulation vector of the Fulde-Ferrell-Larkin-Ovchinnikov state in CeCoIn$_5$\\ revealed by high-resolution magnetostriction measurements}

\author{Shunichiro Kittaka}
\affiliation{Department of Physics, Chuo University, Bunkyo-ku, Tokyo 112-8551, Japan}
\author{Yohei Kono}
\affiliation{Department of Physics, Chuo University, Bunkyo-ku, Tokyo 112-8551, Japan}
\author{Kaito Tsunashima}
\affiliation{Department of Physics, Chuo University, Bunkyo-ku, Tokyo 112-8551, Japan}
\author{Daisuke Kimoto}
\affiliation{Department of Physics, Chuo University, Bunkyo-ku, Tokyo 112-8551, Japan}
\author{Makoto Yokoyama}
\affiliation{Faculty of Science, Ibaraki University, Mito, Ibaraki 310-8512, Japan and Institute of Quantum Beam Science, Ibaraki University, Mito, Ibaraki 310-8512, Japan}
\author{Yusei Shimizu}
\affiliation{Institute for Materials Research (IMR), Tohoku University, Oarai, Ibaraki 311-1313, Japan}
\author{Toshiro Sakakibara}
\affiliation{Institute for Solid State Physics, University of Tokyo, Kashiwa, Chiba 277-8581, Japan}
\author{Minoru Yamashita}
\affiliation{Institute for Solid State Physics, University of Tokyo, Kashiwa, Chiba 277-8581, Japan}
\author{Kazushige Machida}
\affiliation{Department of Physics, Ritsumeikan University, Kusatsu, Shiga 525-8577, Japan}

\date{\today}

\begin{abstract}
The Fulde-Ferrell-Larkin-Ovchinnikov (FFLO) state is an exotic superconducting phase formed by Cooper pairs with finite center-of-mass momentum $\Vec{q}$.
On theoretical grounds, the superconducting order parameter in the FFLO state is spatially modulated along the $\Vec{q}$ vector, and 
the emergence of an associated anisotropy is expected at the phase transition from the Abrikosov state to the FFLO state.
Here, we report the results of high-resolution magnetostriction measurements for a single crystal of CeCoIn$_5$ around $B \parallel c$. 
We find two anomalies in the magnetostriction along the $c$ axis, parallel to the magnetic-field orientation.
In sharp contrast, this $B_{\rm K}$ anomaly disappears in the magnetostriction along the $a$-axis direction, perpendicular to the magnetic-field orientation.  
To explain this uniaxial expansion, we suggest a possibility that 
the FFLO transition occurs slightly below the upper critical field, and 
the FFLO modulation vector parallel to the applied magnetic field gives rise to the anisotropic response.
\end{abstract}

\maketitle

\begin{figure}
\includegraphics[width=3.2in]{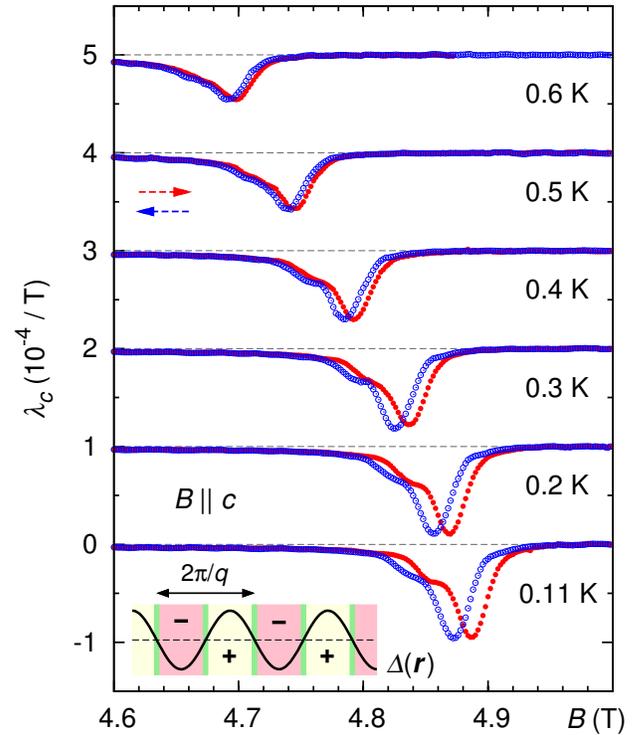}
\caption{
The $c$-axis magnetostriction coefficient $\lambda_c(B)=(\partial L_c/\partial B)/L_c$ at several temperatures for $B \parallel c$.
Each set of data is shifted vertically by $10^{-4}$ T$^{-1}$ for clarity. 
Closed (open) circles represent the data taken in the field-increasing (-decreasing) processes.
Inset shows a schematic image of the spatial modulation of the order parameter $\Delta(\Vec{r})$ in the FFLO state along the modulation vector $\Vec{q}$.
}
\label{Tdep}
\end{figure}

The Fulde-Ferrell-Larkin-Ovchinnikov (FFLO) state \cite{Fulde1964PR,Larkin1964ZETF} is a prototype example of the so-called pair density wave superconductivity (inset of Fig.~\ref{Tdep}) \cite{Agterberg2020ARC,Agterberg2008NP} 
where translational symmetry is spontaneously broken in addition to $U(1)$ gauge symmetry. 
It is a topologically interesting object~\cite{Mizushima2005PRL} and
has been long sought in a wide range of research fields, from condensed matter \cite{Matsuda2007JPSJ} to cold neutral atoms.\cite{Zwierlein2006Sciecne,Partridge2006Science,Mizushima2005PRL-2}
In order to realize the FFLO state, several strict conditions are required to be fulfilled.\cite{Matsuda2007JPSJ}
For example, the system must be very clean because the FFLO state is easy to be broken by a small amount of impurities and/or defects. 
Furthermore, the orbital pair-breaking effect needs to be sufficiently weaker than the Pauli-paramagnetic effect.
Therefore, for a long period of time, there were few experimental reports on the FFLO state.
However, in recent years, there has been a gradual increase in the number of materials that are expected to realize the FFLO state, 
such as quasi-low-dimensional organic superconductors,\cite{Singleton2000JPCM,Uji2001Nature,Uji2006PRL,Shinagawa2007PRL,Yonezawa2008PRL,Yonezawa2008JPSJ,Mayaffre2014NatPhys,Imajo2021PRB,Imajo2022NatCom} iron-based superconductors,\cite{Burger2013PRB,Cho2017PRL,Kasahara2020PRL} and Sr$_2$RuO$_4$.\cite{Kittaka2018JPSJ,Kinjo2022Science}

The heavy-fermion superconductor CeCoIn$_5$ is also a good candidate realizing the FFLO state.
It has been well established as a spin-singlet $d_{x^2-y^2}$-wave superconductor.\cite{Curro2001PRB,Kohori2001PRB,Izawa2001PRL,An2010PRL,Allan2013NatPhys,Zhou2013NatPhys}
In the heavy-fermion system, the orbital pair-breaking effect is sufficiently weak due to the large effective mass of heavy quasiparticles.
Indeed, the emergence of a first-order superconducting transition at the upper critical field $\Bc2$ below 0.7~K, as well as the suppression of $\Bc2$ at low temperatures, have been reported in CeCoIn$_5$,\cite{Bianchi2002PRL,Tayama2002PRB,Murphy2002PRB}
indicating that the Pauli-paramagnetic effect overcomes the orbital pair-breaking effect.
Moreover, a specific-heat anomaly has been found inside the superconducting phase under an in-plane magnetic field.\cite{Bianchi2003PRL,Radovan2003Nature}
This anomaly may be attributed to a transition from the uniform superconducting state, i.e., Abrikosov vortex state, to the FFLO state.\cite{Ichioka2007PRB,Suzuki2020PRB}
The double-peak spectral structure detected slightly below $\Bc2$ from nuclear magnetic resonance (NMR) experiments also supports the presence of a high-field phase.\cite{Kakuyanagi2005PRL,Kumagai2006PRL,Kumagai2009JPCS,Kumagai2011PRL}
Thus, CeCoIn$_5$ has a high-field superconducting phase, particularly for $B \parallel ab$.

However, it remains controversial whether this high-field phase is the FFLO phase.
From NMR and neutron-scattering experiments, it has been revealed that, in the high-field phase for $B \parallel ab$ (the so-called $Q$ phase), 
a spin density wave (SDW) order coexists with superconductivity.\cite{Young2007PRL,Kenzelmann2008Science}
Because the magnetic structure in this $Q$ phase is independent of the direction of the in-plane magnetic field, 
it was suggested that the $Q$ phase is not driven by the FFLO state.\cite{Kenzelmann2010PRL}
Recent theoretical studies and thermal conductivity measurements have also suggested that the FFLO state competes with the SDW phase.\cite{Lin2020PRL}
The coexistence of SDW order and superconductivity complicates the interpretation of the high-field phase in CeCoIn$_5$ for $B \parallel ab$.

Then, it seems reasonable to turn our attention to a possible high-field phase in CeCoIn$_5$ for $B \parallel c$ 
because the magnetic ordering is quickly suppressed by tilting the magnetic field from the $ab$ plane.\cite{Blackburn2010PRL}
Indeed, no spectrum broadening is reported in the recent NMR measurements for $B \parallel c$,\cite{Taniguchi2020JPSCP} 
showing the absence of the magnetic instability around $\Bc2$.
Therefore, if the high-field phase exists in $B \parallel c$, it is expected to be a pure FFLO phase.
From NMR experiments, double-peak structure, similar to the one for $B \parallel ab$,\cite{Kakuyanagi2005PRL,Kumagai2011PRL} has been observed in the range $4.7\ {\rm T} \lesssim B \lesssim \Bc2$ for $B \parallel c$.\cite{Kumagai2006PRL,Kumagai2009JPCS}
However, there is no other crucial evidence for the occurrence of the high-field phase in $B \parallel c$.
In order to provide further experimental evidence for the FFLO transition in $B \parallel c$, 
we have performed high-resolution magnetostriction measurements.

High-quality single crystals of CeCoIn$_5$ were grown by the self-flux method.
This paper focuses on the results of one of three samples used; the others are shown in Secs. I and II of the Supplemental Material (SM).\cite{SM}
The isothermal magnetostriction, $\Delta L_i(B)=L_i(B) - L_i(B_0)$, and thermal expansion, $\Delta L_i(T)=L_i(T) - L_i(T_0)$, along the $i$ ($i=a$ or $c$) axis were measured 
using a homemade capacitance dilatometer in a dilution refrigerator (Kelvinox AST Minisorb, Oxford). 
Here, $B_0$ ($T_0$) represents a selected constant field (temperature).
The sample lengths were 1.34 and 0.9~mm along the $a$ and $c$ axes, respectively. 
The crystalline axes were determined using an x-ray back-scattering Laue camera (RASCO-BL II, Rigaku).
A magnetic field $B$ was generated using a 7-T split-pair magnet in the horizontal $x$ direction.
For $L_c$ ($L_{a}$) measurements, the measurement direction of the compact dilatometer was set to be parallel to the $x$ ($z$) axis, 
so that the magnetic field can be applied along the $c$ axis.
The field-angle $\phi$ dependences of $L_c$ and $L_{a}$ were investigated 
by rotating the sample together with the refrigerator around the vertical $z$ direction using a stepper motor at the top of the magnet Dewar,
where $\phi$ denotes the field angle measured from the $c$ axis to the $b$ axis. 
Although, due to the non-rectangular shape of the sample, the $a$ axis was not perfectly aligned to the measurement direction of $L_a$, 
this misalignment does not affect the conclusion of this paper~(see Sec. IV of SM \cite{SM} for more details).

Figure \ref{Tdep} presents the field derivative data, $\lambda_c(B)=(\partial L_c/\partial B)/L_c$ at several temperatures in the field-increasing and -decreasing processes.
As shown in Fig.~\ref{Tdep}, $|\lambda_c(B)|$ shows a sharp peak at $\Bc2$ with a hysteresis that develops below 0.6~K, 
demonstrating the first-order superconducting transition at $\Bc2$.\cite{Bianchi2002PRL,Tayama2002PRB}
The sharp peak of $|\lambda_c(B)|$ at $\Bc2$ demonstrates the high quality of the present sample.
Most remarkably, slightly below $\Bc2$, a kink anomaly appears at $B_{\rm K}$ concomitantly with the development of the hysteresis at low temperatures. 
We confirm the reproducibility of this $B_{\rm K}$ anomaly in the measurement of another sample (see Sec. II of SM \cite{SM}).
Indeed, the specific-heat anomaly close to $\Bc2$ is suggested in the previous report.\cite{Bianchi2003PRL}
It should be noted that our high-resolution magnetostriction measurements performed at small steps of $\sim 0.002$~T near $\Bc2$ allow us to clearly detect 
the $B_{\rm K}$ anomaly in $B \parallel c$ that was not reported in the previous work.\cite{Correa2007PRL,Takeuchi2002JPCM}

\begin{figure}
\includegraphics[width=3.2in]{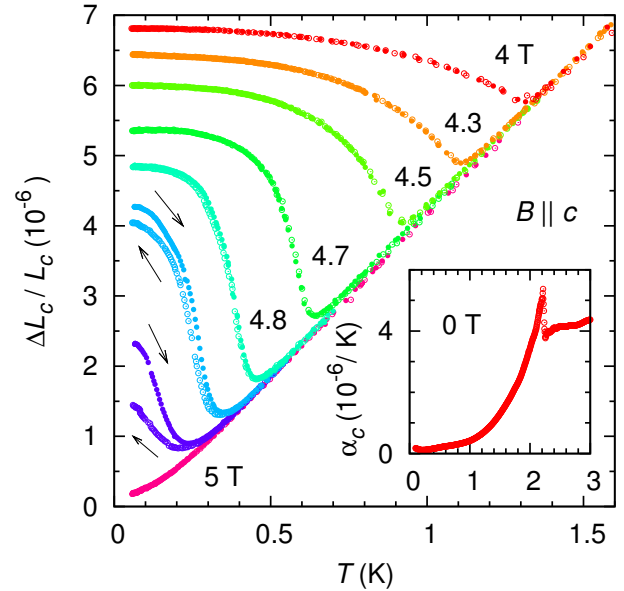}
\caption{
Thermal expansion $\Delta L_c/L_c$ measured at 5, 4.89, 4.85, 4.8, 4.7, 4.5, 4.3, and 4~T (from bottom to top) for $B \parallel c$.
Closed and open circles represent the data taken in the warming and cooling processes, respectively.
Inset shows temperature dependence of the thermal expansion coefficient $\alpha_c=(\partial L_c/\partial T)/L_c$ along the $c$ axis at 0~T. 
}
\label{thermal}
\end{figure}

The inset of Fig.~\ref{thermal} displays the linear thermal expansion coefficient $\alpha_c$ in zero field, 
where $\alpha_c$ denotes $(\partial L_c/\partial T)/L_c$.
A sharp anomaly is seen in $\alpha_c$ at $\Tc=2.25$~K in zero field; 
$\alpha_c$ exhibits no sign of $\Tc$ distribution within the resolution limit. 
Figure \ref{thermal} shows the thermal expansion $\Delta L_c(T)/L_c$ at several magnetic fields for $B \parallel c$.
The hysteresis behavior in $\Delta L_c(T)$ becomes prominent above 4.8~T in the superconducting state,
while no difference was found between cooling and warming in the $\Delta L_c(T)$ data at 4.7~T.
The $B_{\rm K}$ anomaly is less clearly detected from the thermal expansion (temperature scan)
because the superconducting transition at $T_{\rm c}$ affects the thermal expansion in wide temperature range (see Sec. III of SM~\cite{SM}).

\begin{figure}
\includegraphics[width=3.2in]{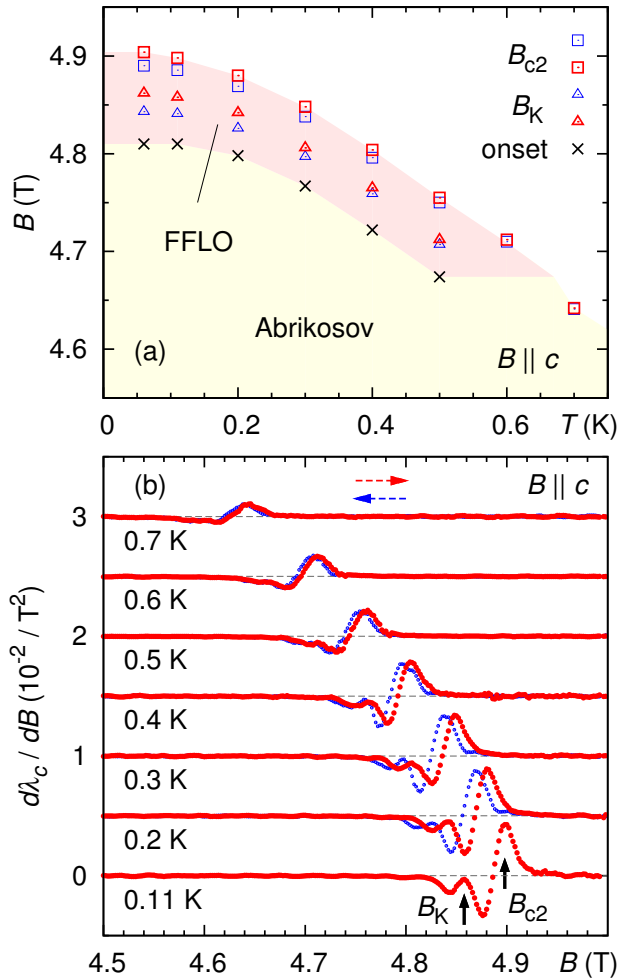}
\caption{
(a) Field-temperature phase diagram of CeCoIn$_5$ for $B \parallel c$.
(b) Field dependence of $d\lambda_{c}/dB$ at several temperatures.
Both increasing and decreasing field data are shown above 0.2~K (only the former at 0.11 K).
Each set of data in (b) is shifted vertically by $5\times10^{-3}$ T$^{-2}$ for clarity. 
The positions of $\Bc2$ [squares in (a)] and $B_{\rm K}$ (triangles) are determined by large and small peaks in $d\lambda_c/dB$, respectively, as indicated by solid arrows in (b). 
The symbols shown in red (blue) are determined from the field-increasing (-decreasing) data.
Crosses represent the field above which $\lambda_{c}(B)$ starts to change markedly toward $B_{\rm K}$ with increasing $B$, 
possibly corresponding to a boundary between the Abrikosov state and the FFLO state.
}
\label{HT}
\end{figure}

Figure \ref{HT}(a) shows the field-temperature phase diagram of CeCoIn$_5$ for $B \parallel c$, 
in which the positions of $\Bc2$ and $B_{\rm K}$ are determined by the two peaks in $d\lambda_c/dB$ as indicated by arrows in Fig.~\ref{HT}(b).
The $B_{\rm K}$ anomaly becomes indistinguishable above 0.6 K due to the broadening of the $\Bc2$ transition.
The tricritical point between the FFLO, homogeneous Abrikosov vortex, and paramagnetic normal states can be determined by the onset critical field 
above which the superconducting transition becomes first order. 
To determine this critical magnetic field, we estimate the temperature dependence of the magnetic hysteresis at $\Bc2$ between the field increasing and decreasing measurements (Fig.~\ref{HT}). 
The magnetic hysteresis between these measurements appears at around 0.6 K and 4.7 T for $B \parallel c$, 
which is suggested as the tricritical point (see also Fig. S4 of SM~\cite{SM}).

\begin{figure}
\includegraphics[width=3.2in]{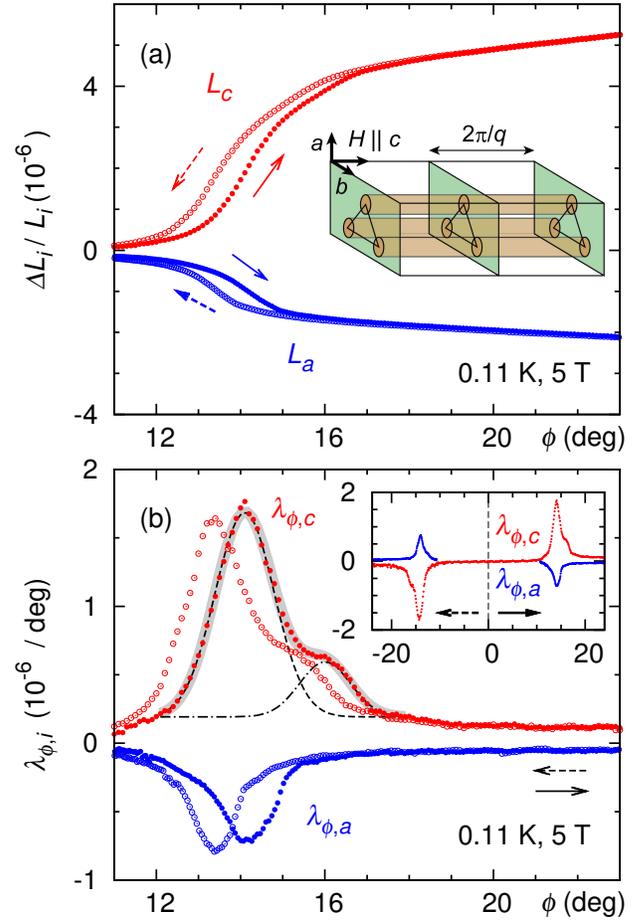}
\caption{
Field angle $\phi$ dependence of (a) $\Delta L_i/L_i$ at 0.11 K and 5~T, 
and (b) their field-angle derivatives, $\lambda_{\phi,i}=(\partial L_i/\partial\phi)/L_i$ ($i=a$ or $c$).
Closed (open) symbols represent the data taken in the $\phi$-increasing (decreasing) process.
The gray solid line in (b) represents the simulation spectrum consisting of two Gaussian functions (dashed and dash-dotted lines).
Inset in (a) shows a schematic view of the flux lines and the FFLO modulation. 
The periodic nodal planes, separated by $L=2\pi/q$, run perpendicular to the flux lines that form a vortex lattice in the $ab$ plane.
Inset in (b) shows a wider view of $\lambda_{\phi,i}(\phi)$ in the $\phi$-increasing (decreasing) process for $\phi>0$ ($\phi<0$).
}
\label{phi}
\end{figure}

Due to the anisotropy in $\Bc2$ of CeCoIn$_5$ ($\Bc2^{\parallel ab}/\Bc2^{\parallel c}\sim 2.4$), 
the superconducting transition can be induced by rotating a magnetic field from the $c$ axis 
when $\Bc2^{\parallel c}<B<\Bc2^{\parallel ab}$.
Figure \ref{phi}(a) shows the field-angle $\phi$ dependence of $\Delta L_c(\phi)=L_c(\phi)-L_c(\phi_0)$ measured at 0.11~K under a magnetic field of 5~T rotated within the $bc$ plane.
Here, $\phi_0$ is a selected constant field angle.
A first-order superconducting transition with clear hysteresis has been observed in $\Delta L_c(\phi)$ at $\phi \sim 14^\circ$.
The field-angle derivative data, $\lambda_{\phi,c}(\phi)=(\partial L_c/\partial \phi)/L_c$, are presented in Fig.~\ref{phi}(b).
In $\lambda_{\phi,c}(\phi)$, the $B_{\rm K}$ anomaly can be seen at $\phi \sim 16^\circ$,
which is compatible with the one observed in $\lambda_c(B)$ (Fig.~\ref{Tdep}). 
The height of the $B_{\rm K}$ anomaly does not depend on the direction of the field angle sweep,
whereas the $\Bc2$ anomaly at $\phi\sim14^\circ$ 
is more prominent for the transition from normal to superconducting state [closed symbols in Fig.~\ref{phi}(b)] 
than for the transition from superconducting to normal state [open symbols in Fig.~\ref{phi}(b)].
Thus, two anomalies exhibit qualitatively different features, likely stemming from different origins.

Furthermore, the $a$-axis magnetostriction $\Delta L_a(\phi)=L_a(\phi)-L_a(\phi_0)$ of the same sample was measured 
under a rotating magnetic field within the plane normal to the measurement direction, i.e., the approximate $bc$ plane.
Figures~\ref{phi}(a) and \ref{phi}(b) show the $\phi$ dependences of $\Delta L_a(\phi)$ and its angle derivative $\lambda_{\phi,a}(\phi)=(\partial L_a/\partial \phi)/L_a$, respectively.
The change in $\Delta L_a/L_a$ at $\Bc2$ is $2\times 10^{-6}$, about half of the change in $\Delta L_c/L_c$. 
These results are consistent with the previous report.\cite{Takeuchi2002JPCM,Correa2007PRL}
In sharp contrast to $\lambda_{\phi,c}(\phi)$, 
$\lambda_{\phi, a}(\phi)$ shows one sharp transition without the second anomaly at $B_{\rm K}$.
The observed peak in $|\lambda_{\phi, a}(\phi)|$ has similar features to the main peak in  $|\lambda_{\phi,c}|$ (see Sec. IV of SM \cite{SM}).

As shown in the inset of Fig. \ref{phi}(b), both $\lambda_{\phi,c}$ and $\lambda_{\phi,a}$ are symmetric with respect to the angle direction. 
This symmetric angle dependence in $\lambda_{\phi,c}$ eliminates the possibility that the $B_{\rm K}$ anomaly is caused by a domain with a tilted $c$ axis in the sample, 
because such a domain should show the $B_{\rm K}$ anomaly at a smaller or larger field angle
when the magnetic field is rotated in the other direction.
The results of the Gaussian fits to $\lambda_{\phi,c}$ [a solid line in Fig.~\ref{phi}(b)] reveal that 
a full width at half maximum (FWHM) of the $B_{\rm K}$ anomaly ($\sim 1.42^\circ\pm0.05^\circ$) [a dash-dotted line in Fig.~\ref{phi}(b)] is narrower than 
the FWHM of the $B_{\rm c2}$ anomaly ($\sim 1.65^\circ\pm0.02$) [a dashed line in Fig.~\ref{phi}(b)] (see Sec. IV of SM \cite{SM} for more details).
From this fact, the $B_{\rm K}$ anomaly is unlikely to be caused by sample inhomogeneities that exist in regions not detected from the $L_a$ measurements 
because suppression of $\Bc2$ by impurities and/or defects usually results in a wider distribution of $\Bc2$.

Let us discuss why the $B_{\rm K}$ anomaly exists only in $L_c$ when $B \parallel c$.
The most plausible origin is the formation of the FFLO state with $\Vec{q} \parallel B$, as detailed below.

The FFLO state is characterized by a periodic spatial modulation of the superconducting order parameter
$\Delta(\Vec{r})$ with the wave vector $\Vec{q}$.
The direction of the $\Vec{q}$ vector is determined by the combination of the relative stabilities 
between the vortex lattice configuration  and the nesting condition, where the Zeeman-split Fermi surfaces are maximally touched under the translation by $\Vec{q}$. 
When the FFLO state is realized in a superconductor with an isotropic spherical Fermi surface, 
the $\Vec{q}$ vector directs parallel to the field direction; $\Vec{q} \parallel \Vec{B}$.
This is because the vortex lattice configuration is least perturbed by the formation of the FFLO state, 
otherwise the two factors of the relative stabilities, i.e., vortex lattice configuration and nesting condition, interfere each other [see the inset of Fig.~\ref{phi}(a)]. 
For CeCoIn$_5$, it is necessary to consider the nesting condition based on the actual band structure.\cite{Shishido2003JPCM,Settai2001JPCM,Hall2001PRB}
The main Fermi surfaces of CeCoIn$_5$ are the two heavy electron bands ($\alpha$ and $\beta$ bands) with the warped cylindrical shape, 
open along the $c$ axis at the four corners of the tetragonal Brillouin zone.
For $B \parallel c$, the optimal $\Vec{q}$ direction is parallel to the $c$ axis under this uniaxial symmetry situation 
because the two Fermi surfaces are nested circularly around the warped neck region, 
as confirmed by a model calculation.\cite{Shimahara2021JPSJ}

\begin{figure}
\includegraphics[width=3.2in]{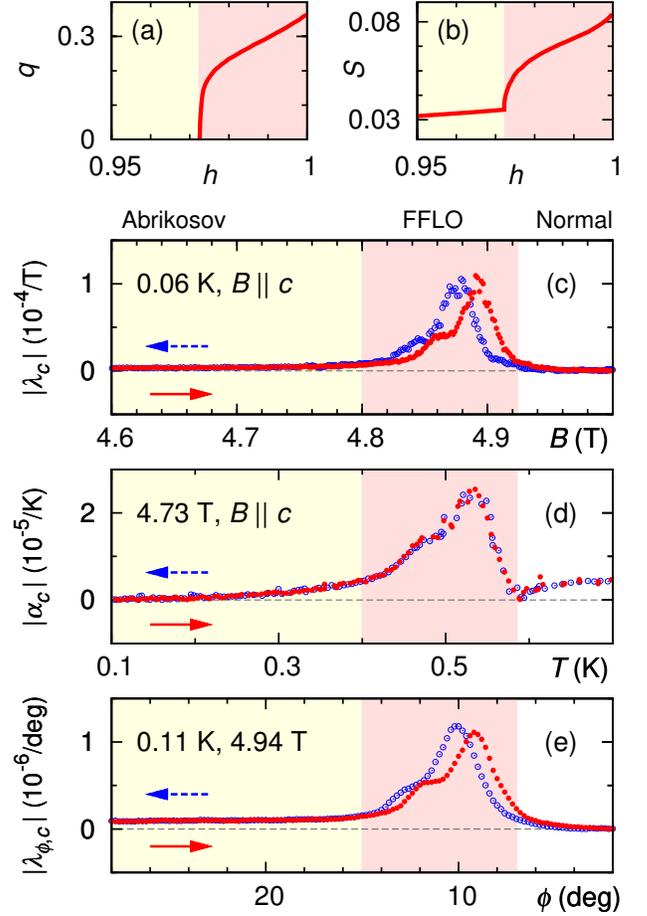}
\caption{
(a) The wave number $q$ normalized by the coherent length and 
(b) the normalized entropy $S=S_{\rm s}(T)/S_{\rm n}(\Tc)$ as a function of $h=H/\Hc2$, taken from Ref.~\onlinecite{Suzuki2020PRB}.
Here, $S_{\rm s}$ ($S_{\rm n}$) is the entropy in the superconducting (normal) state.
(c) $|\lambda_c|$ as a function of $B$ at $T=0.06$~K for $B \parallel c$.
(d) $|\alpha_c|$ as a function of $T$ at $B=4.73$~T for $B \parallel c$.
(e) $|\lambda_{\phi,c}|$ as a function of the angle $\phi$ at $B=4.94$~T and $T=0.11$~K.
The yellow and the pink colored regions represent the Abrikosov state and the FFLO state, respectively; 
the boundary in $|\lambda_c|$ and $|\lambda_{\phi,c}|$ ($|\alpha_c|$) is determined by the onset of the $B_{\rm K}$ anomaly [inferred from the phase diagram in Fig.~\ref{HT}(a)].
}
\label{f0}
\end{figure}

In the FFLO state, $\Delta(\Vec{r})$ has the nodal planes perpendicular to the $\Vec{q}$ direction whose separation is given by $L=2\pi/q$, where $q=|\Vec{q}|$.
Thus, the flux lines intersect perpendicular to the nodal planes, as shown in the inset of Fig.~\ref{phi}(a).
The nodal plane accommodates the excess imbalanced spin component produced by the applied magnetic field through the Pauli paramagnetic effect 
because the sign change of $\Delta(\Vec{r})$ allows the Andreev bound state at the zero energy position.\cite{mizu} 
The nodal plane or domain wall contains the excess paramagnetic moment, accompanying the longitudinal strain wave.\cite{Ichioka2007PRB,Suzuki2020PRB}

Upon entering the FFLO state from the Abrikosov vortex state with increasing $B$, 
the wave number $q$ quickly increases from $q=0$ at the transition field as shown in Fig. \ref{f0}(a). 
This abrupt increase is a universal feature, independent of theoretical frameworks.\cite{Machida1984PRB}
This rapid growth of $q$ results in an upward field dependence common to other physical quantities such as the entropy [Fig. \ref{f0}(b)], 
the density of state, and the paramagnetic moment.
Therefore, $L_c$ is expected to show a similar dependence on $B$, $T$ and $\phi$ near the FFLO transition.
Indeed, field, temperature, and field-angle derivatives of the change in $\Delta L_c/L_c$, corresponding to $\lambda_c$, the linear thermal expansion coefficient $\alpha_c$, and $\lambda_{\phi,c}$ respectively, 
exhibit quick change at the Abrikosov-FFLO boundary along the three different paths ($B$, $T$ and $\phi$), 
as shown in Figs.~\ref{f0}(c)-\ref{f0}(e). 
In this phase transition, the phase boundary between the Abrikosov and FFLO states corresponds to the lower onset field as shown in Fig.~\ref{f0}(b). 
Therefore, the phase boundary is determined by the lower onset of the $B_{\rm K}$ anomaly [crosses in Fig. \ref{HT}(a)], rather than its peak position.

The absence of the $B_{\rm K}$ anomaly in the $L_a$ measurements is also consistent with the FFLO scenario, 
because the spatial modulation brought by the FFLO transition runs only along $\Vec{q} \parallel c$ with keeping the uniformity in the $ab$ plane [see the inset of Fig.~\ref{phi}(a)].
Indeed, even if the $a$-axis length is maximally distorted according to the Poisson's ratio ($0.3$ for typical metals), the $B_{\rm K}$ anomaly in $|\lambda_{\phi,a}|$ would be at most $0.15\times10^{-6}$/deg,
which should be less prominent than the one observed in $|\lambda_{\phi,c}|$.
This anisotropic magnetostriction response of the $B_{\rm K}$ anomaly is reminiscent of a FFLO transition suggested in $\kappa$-(BEDT-TTF)$_2$Cu(NCS)$_2$,\cite{Imajo2022NatCom}
whose acoustic response depends on the sound propagation direction. 
Moreover, the observed field range $\Delta B_{\rm LO}$, which is estimated by the width of the magnetic field area between the onset of the $B_{\rm K}$ anomaly and $\Bc2$ [the pink region in Fig. \ref{f0}(c)], relative to $\Bc2$, $\Delta B_{\rm LO} / \Bc2 \sim 2.4$\%, 
is comparable to the theoretical calculation ($\sim 2.7\%$).\cite{Suzuki2020PRB}
Thus, the $B_{\rm K}$ anomaly can be well understood if the FFLO state with $\Vec{q} \parallel c$ is formed in CeCoIn$_5$ for $B \parallel c$, 
which should be scrutinized by further measurements in future to directly detect the spatial modulation of the superconducting gap structure.

In summary, we have performed high-resolution magnetostriction measurements on CeCoIn$_5$ along the $c$ and $a$ axes 
in the magnetic field applied around the $c$ axis.
We find a double superconducting transition at $B_{\rm K}$ and $\Bc2$ in the $c$-axis magnetostriction.
On the other hand, this $B_{\rm K}$ anomaly is absent in the $a$-axis magnetostriction of the same sample.
To explain this anisotropic expansion, we suggest a possibility that
the anisotropic length changes of $L_c$ and $L_a$ are a direct manifestation of the FFLO formation, and 
that the modulation vector $\Vec{q}$ points parallel to the field direction along the $c$ axis.
The characteristic changes in the $|\lambda_c(B)|$ curves with an upward curvature are consistent with the theoretical prediction assuming the FFLO formation.
This study paves the way for determining the $\Vec{q}$-vector orientation of the FFLO state 
from the magnetostriction in different measurement directions.

\begin{acknowledgments}
We thank Kitami Tsuji and Hirohiko Sato for their support.
A part of this work was carried out as joint research in ISSP. 
This work was also supported by KAKENHI (JP20K20893, JP23H04868, JP17K05553, JP21K03455, JP17K05529, JP20K03852, JP23H01116) from JSPS and Chuo University Grant for Special Research.
\end{acknowledgments}

\clearpage
\onecolumngrid
\appendix

\begin{center}
{\large Supplemental Material for \\
Modulation vector of the Fulde-Ferrell-Larkin-Ovchinnikov state in CeCoIn$_5$ \\revealed by high-resolution magnetostriction measurements\\}
\vspace{0.1in}
Shunichiro Kittaka$^{1}$, Yohei Kono$^{1}$, Kaito Tsunashima$^{1}$, Daisuke Kimoto$^{1}$, Makoto Yokoyama$^{2}$, Yusei Shimizu$^{3}$,  \\Toshiro Sakakibara$^{4}$, Minoru Yamashita$^{4}$, and Kazushige Machida$^{5}$\\
{\small 
\textit{$^1$Department of Physics, Chuo University, Bunkyo-ku, Tokyo 112-8551, Japan}\\
\textit{$^2$Faculty of Science, Ibaraki University, Mito, Ibaraki 310-8512, Japan and Institute of Quantum Beam Science, Ibaraki University, Mito, Ibaraki 310-8512, Japan}\\
\textit{$^3$Institute for Materials Research (IMR), Tohoku University, Oarai, Ibaraki 311-1313, Japan}\\
\textit{$^4$Institute for Solid State Physics, University of Tokyo, Kashiwa, Chiba 277-8581, Japan}\\
\textit{$^5$Department of Physics, Ritsumeikan University, Kusatsu, Shiga 525-8577, Japan}\\
}
(Dated: \today)
\end{center}

\setcounter{figure}{0}
\setcounter{table}{0}
\renewcommand{\thefigure}{S\arabic{figure}}
\renewcommand{\thetable}{S\arabic{table}}

\section*{I. Experimental method}
In this magnetostriction study, we used three samples:
two samples (one sample) grown at Ibaraki (Chuo) University by the self-flux method, which are referred to as the samples s1 and s2 (sample s3).
The thicknesses of the samples s1, s2, and s3 along the $c$ axis are 0.50, 0.51, and 0.90 mm, respectively.
The main text reports the results obtained by using the sample s3.
We developed three homemade capacitance dilatometers with a resolution better than 1~pm.
The $L_c$ measurements on the samples s1 and s2 (sample s3) were performed at University of Tokyo (Chuo University) 
in Oxford dilution refrigerator Kelvinox 100 and 25 (AST minisorb), respectively, 
using a standard (first-developed compact) dilatometer.
The isothermal magnetostriction roughly parallel to the $a$ axis $\Delta L_{a}(B)=L_{a}(B) - L_{a}(B_0)$, 
for the sample s3 was measured using a second-developed compact dilatometer whose resolution is slightly higher than the first one.
The frame diameter and height of a standard (compact) dilatometer are approximately 25 (20) and 45 (25) mm, respectively.
For magnetostriction measurements on the samples s1 and s2 (sample s3), a magnetic field $B$ was generated using 15-T and 9-T solenoid magnets (a 7-T split-pair magnet) in the vertical $z$ (horizontal $x$) direction, respectively.

\section*{II. Sample dependence of the $c$-axis magnetostriction}

\begin{figure}
\includegraphics[width=5.5in]{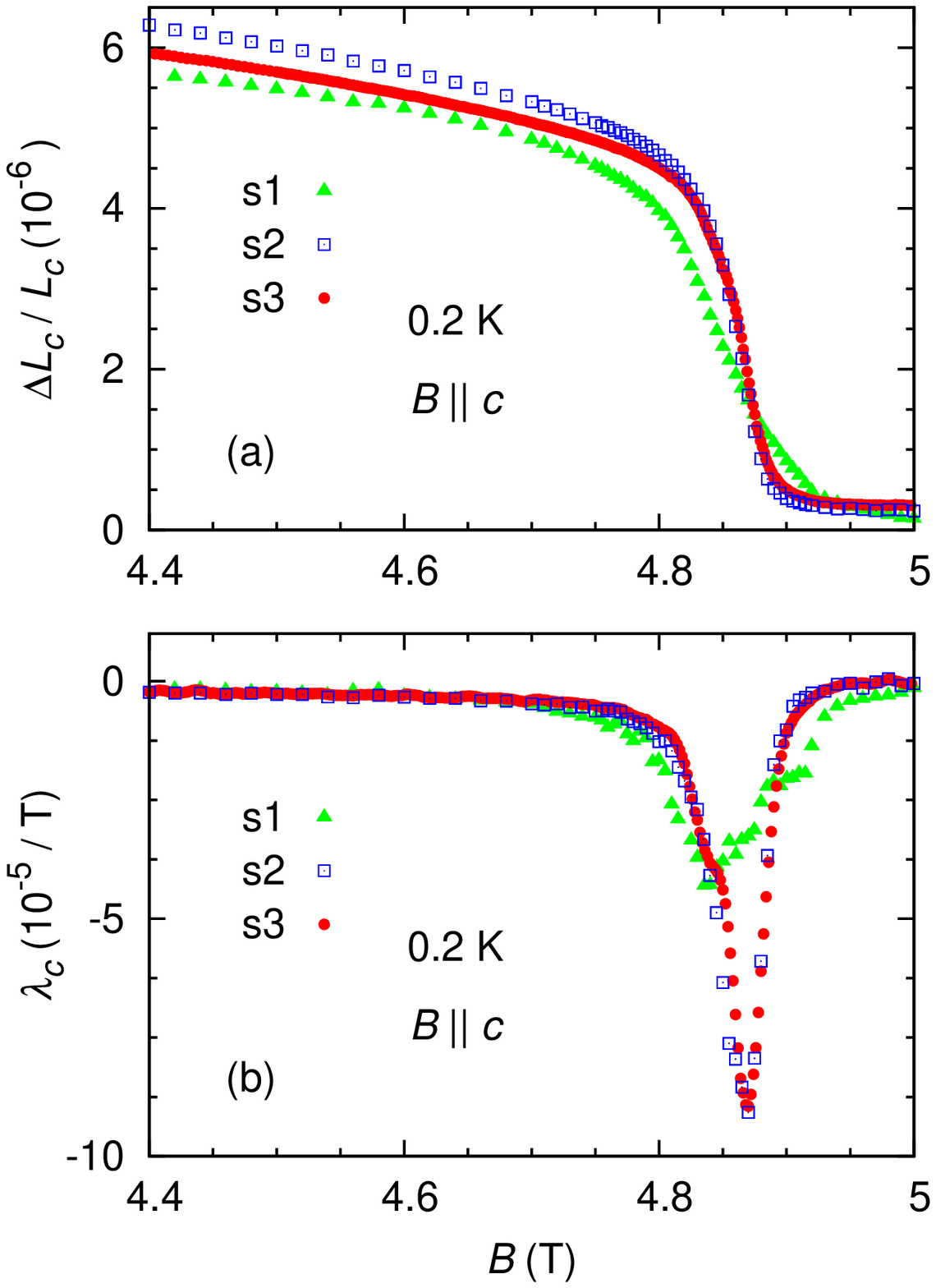}
\caption{
Field dependences of (a) the magnetostriction along the $c$ axis, $\Delta L_c/L_c$, of the samples s1, s2, and s3 on increasing-field process at 0.2~K for $B \parallel c$ and (b) their field derivatives $\lambda_c=(\partial L_c/\partial B)/L_c$.
}
\label{com}
\end{figure}

Figure \ref{com}(a) compares a change in $\Delta L_c(B)/L_c$ of the samples s1, s2, and s3 at 0.2 K for $B \parallel c$ in the field-increasing process. 
In the field range $4.4\ {\rm T}\le B \le 5$~T, $\Delta L_c(B)$ of all samples decreases with increasing $B$ in the superconducting state.
The change of $\Delta L_c/L_c \sim 5 \times 10^{-6}$ in the vicinity of $\Bc2$ is consistent with the previous reports.\cite{Correa2007PRL2,Takeuchi2002JPCM2}
These results reinforce the reliability of our measurements.
A sharp drop at $\Bc2$ in $\Delta L_c(B)$ for the samples s2 and s3 is a sign of the occurrence of a first-order superconducting transition.
By contrast, $\Delta L_c$ of the sample s1 decreases gradually even near $\Bc2$.
These results indicate that the samples s2 and s3 are in higher quality than the sample s1.
The field derivative data of Fig.~\ref{com}(a), $\lambda_c(B)=(\partial L_c/\partial B)/L_c$, are shown in Fig. \ref{com}(b).
The broad anomaly for the sample s1 can be attributed to wide $\Bc2$ distribution originating from sample inhomogeneity.
A relatively sharp dip can be seen at $\Bc2$ in $\lambda_c(B)$ of the samples s2 and s3.

\begin{figure}
\includegraphics[width=5.5in]{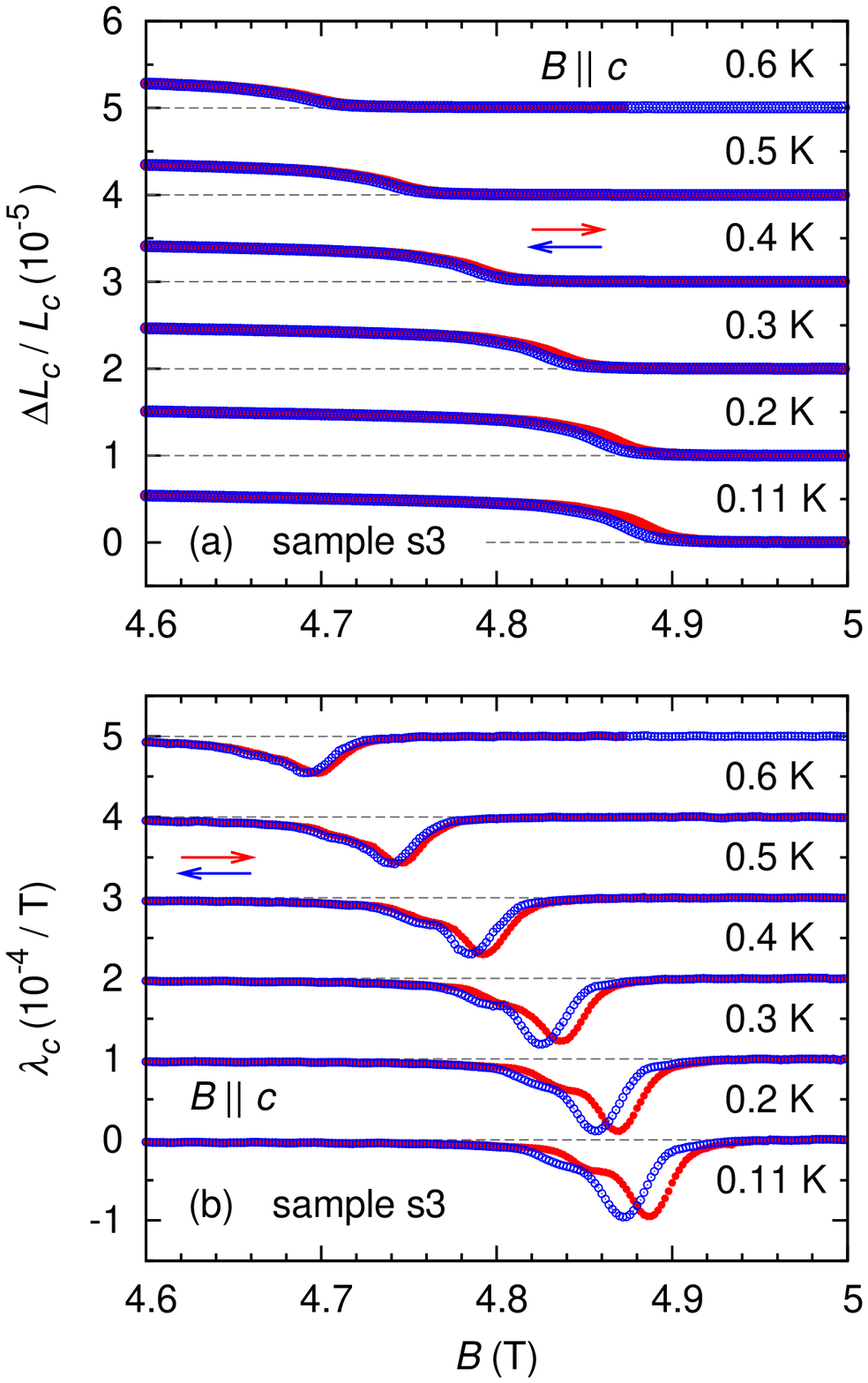}
\caption{
Field dependences of (a) $\Delta L_c/L_c$ and (b) $\lambda_c$ of the samples s3 for $B \parallel c$ at several temperatures.
Each set of data in (a) [(b)] is shifted vertically by $1 \times 10^{-5}$ ($1 \times 10^{-4}$/T) for clarity. 
}
\label{s3}
\end{figure}

\begin{figure}
\includegraphics[width=5.5in]{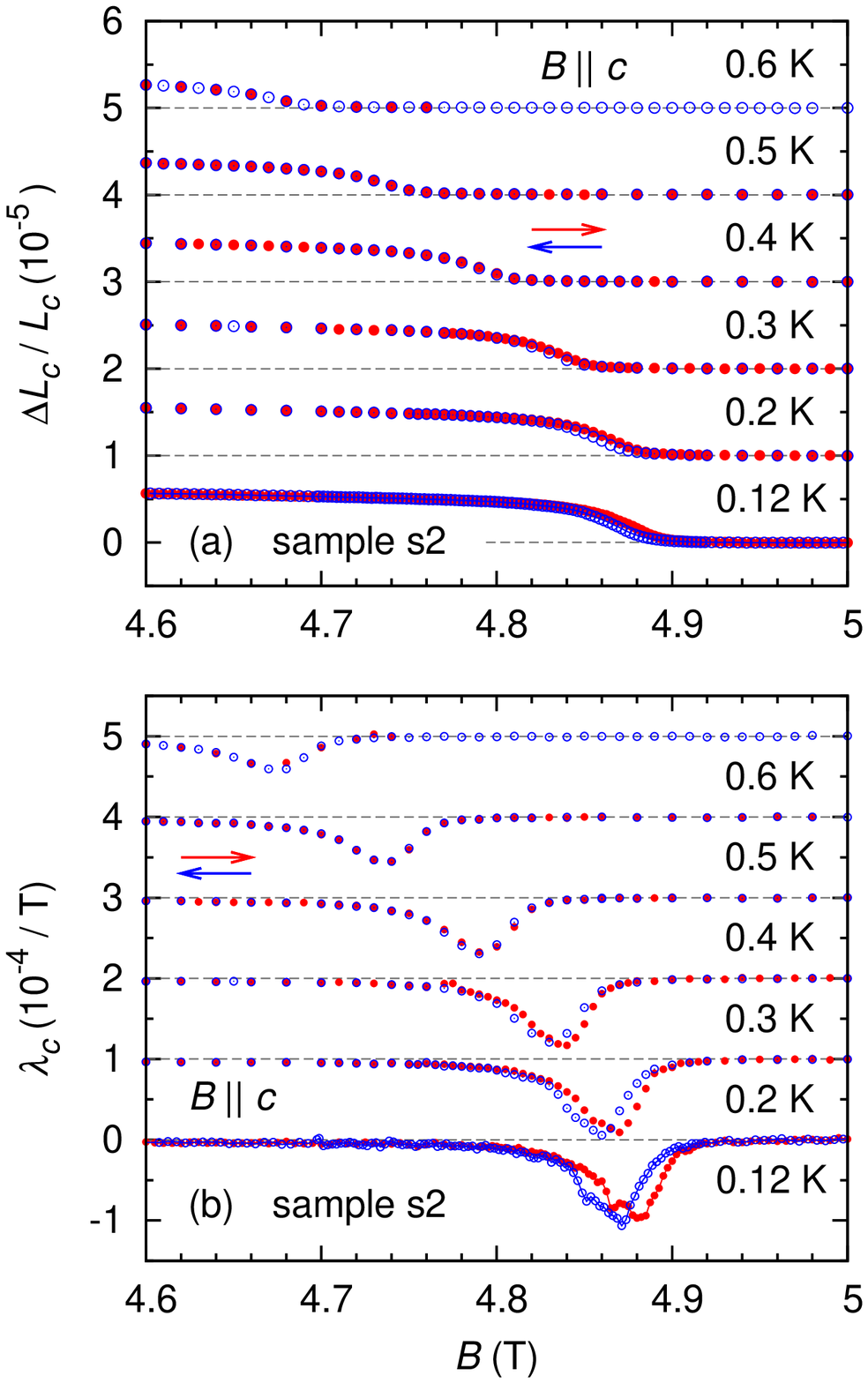}
\caption{
Field dependences of (a) $\Delta L_c/L_c$ and (b) $\lambda_c$ of the samples s2 for $B \parallel c$ at several temperatures.
Each set of data in (a) [(b)] is shifted vertically by $1 \times 10^{-5}$ ($1 \times 10^{-4}$/T) for clarity. 
}
\label{s2}
\end{figure}

\begin{figure}
\includegraphics[width=5.5in]{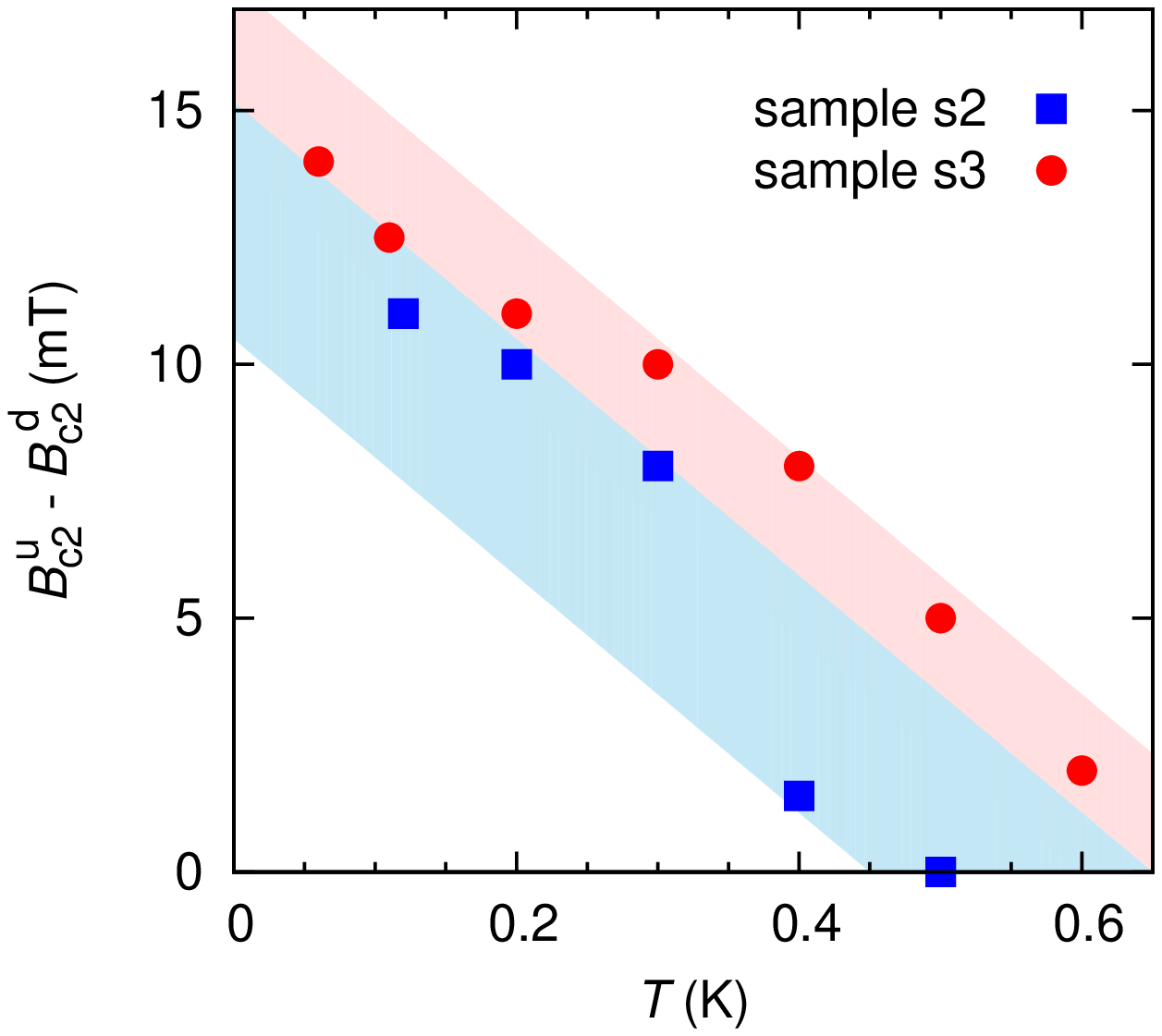}
\caption{
Temperature dependence of the difference between $\Bc2^{\rm u}$ and $\Bc2^{\rm d}$,
where $\Bc2^{\rm u}$ and $\Bc2^{\rm d}$ are $\Bc2$ in the field-increasing and decreasing processes, respectively.
$\Bc2$ of the sample s2 (s3) is defined from the high-field peak in $|\lambda_c(B)|$ [$|d\lambda_c/dB|$].
The shaded areas are guides to the eye.
}
\label{DBc2}
\end{figure}

Figures \ref{s3} and \ref{s2} show the magnetostriction [(a)] and its field derivative [(b)] for the samples s3 and s2, respectively, in both field-increasing and decreasing processes at several temperatures. 
Hysteresis behavior can be seen roughly below 0.6 (0.4)~K for the sample s3 (s2), and the results of these two samples are in reasonable agreements (Fig.~\ref{DBc2}).
A tricritical point between the FFLO, homogeneous Abrikosov vortex, and paramagnetic normal states is suggested to be present around 0.6~K and 4.7~T 
because the superconducting transition at $\Bc2$ becomes a first-order transition 
when the temperature (magnetic field) is below (above) the tricritical point.
A double superconducting transition can be seen at the lowest temperature of roughly 0.1~K for both samples.
The second anomaly at $B_{\rm K}$ slightly below $\Bc2$ can be seen in the wider temperature range for the sample s3 than for the sample s2;
the $B_{\rm K}$ anomaly seems to depend on the sample quality.

\section*{III. Additional thermal expansion data}

\begin{figure}
\includegraphics[width=5.5in]{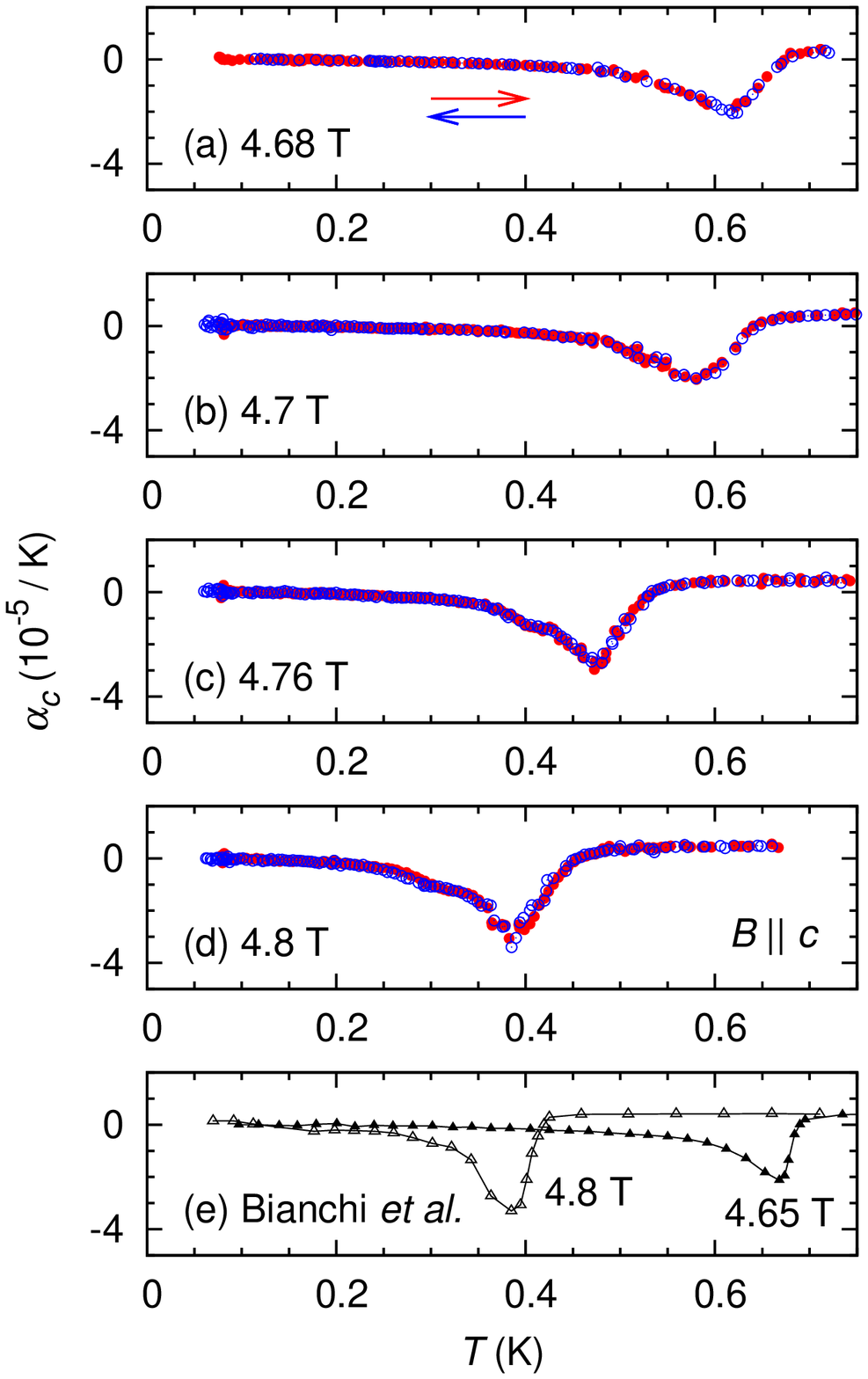}
\caption{
(a)-(d) Temperature dependence of the thermal expansion coefficient, $\alpha_c=(\partial L_c/\partial T)/L_c$, at various magnetic fields along the $c$ axis.
(e) The data taken from the previous report~\cite{Bianchi2002PRL2}.
}
\label{Tdepcom}
\end{figure}

Figure \ref{Tdepcom} shows the thermal expansion coefficient $\alpha_c=(\partial L_c/\partial T)/L_c$ at several magnetic fields for $B \parallel c$.
The $B_{\rm K}$ anomaly is less clearly detected by thermal expansion measurements than by magnetostriction measurements. 
This is caused by the broadening of the superconducting transition in the temperature scan owing to the small slope of the $\Bc2(T)$ phase boundary, $d\Bc2/dT$, at low temperatures. 
Indeed, the $B_{\rm K}$ anomaly is only weakly detected from our thermal expansion measurements in $4.73\ {\rm T} \lesssim B \lesssim 4.8$~T [see Figs. 5(d), \ref{Tdepcom}(c), and \ref{Tdepcom}(d)], 
and is not well resolved at the magnetic fields in the previous work [Fig. \ref{Tdepcom}(e)]. 

\section*{IV. Field-angle dependence of the change in the sample length}

\begin{figure}
\includegraphics[width=5.5in]{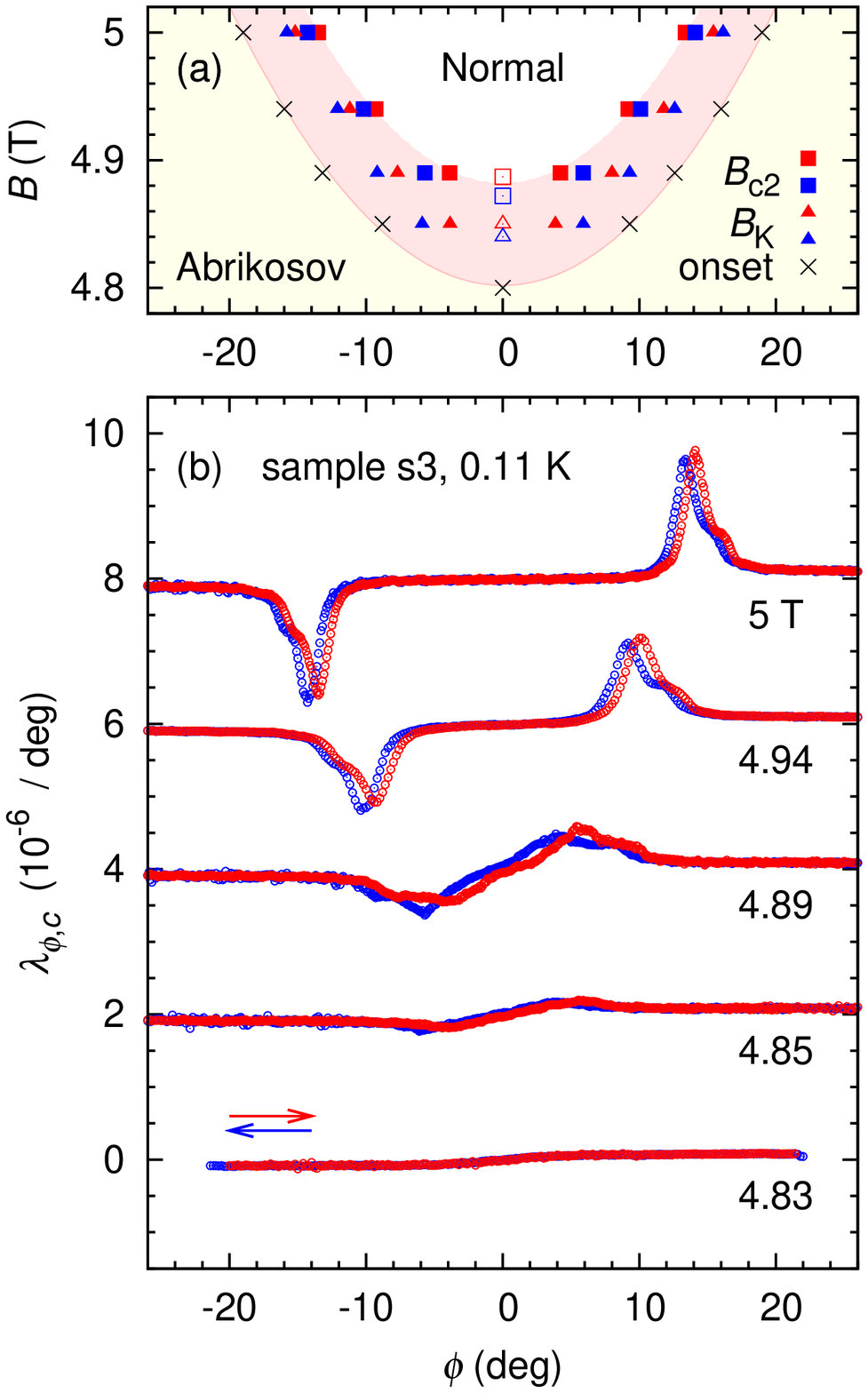}
\caption{
(a) $B$--$\phi$ phase diagram of CeCoIn$_5$. (b) Field-angle $\phi$ dependence of $\lambda_{\phi,c}$ of the sample s3 at 0.11~K in several magnetic fields rotated within the $bc$ plane.
Closed (open) symbols in (a) represent anomalies in (b) [$\lambda_c(B)$ at 0.11 K and $\phi=0$].
Crosses indicate an onset of the $\Delta L_c$ anomaly.
Each set of data in (b) is shifted vertically by $2\times10^{-6}$/deg for clarity. 
}
\label{phic}
\end{figure}

Figure \ref{phic}(b) represents the field-angle $\phi$ dependence of the field-angle derivative $\Delta L_c$ data, 
i.e., $\lambda_{\phi,c}=(\partial L_c/\partial \phi)/L_c$, for the sample s3 at 0.11~K in the field range $4.83~{\rm T}\le B \le 5~{\rm T}$.
Here, $\phi$ denotes the field angle measured from the $c$ axis, and
the magnetic field is rotated within the $bc$ plane.
Above 4.89~T, the two anomalies are clearly observed in $\lambda_{\phi,c}(\phi)$, and the transition field angle becomes larger with increasing magnetic field, as shown in Fig.~\ref{phic}(a).

\begin{figure}
\includegraphics[width=5.5in]{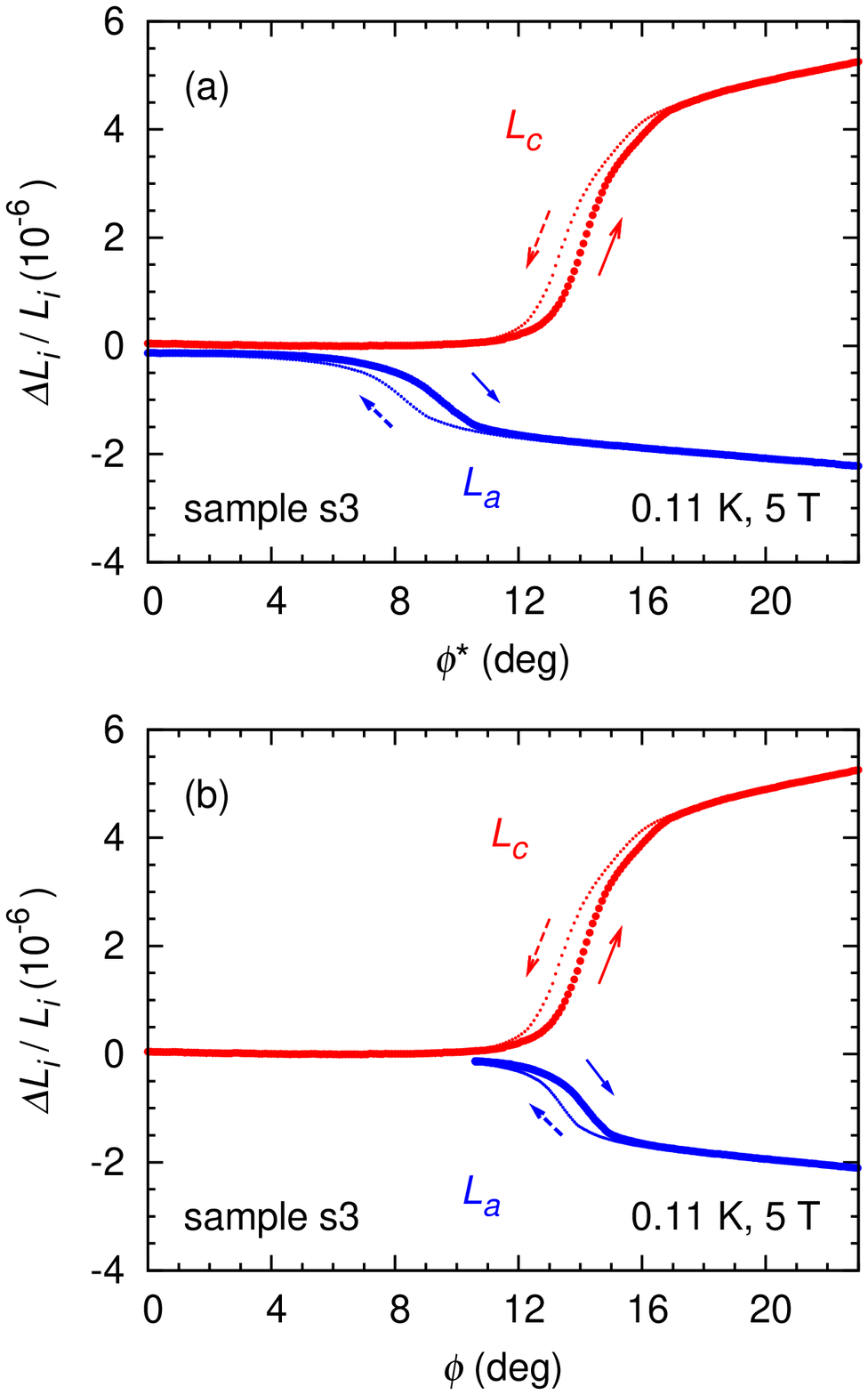}
\caption{
(a) Field-angle $\phi^\ast$ dependence of $\Delta L_i/L_i$ ($i=a$ or $c$) of the sample s3 at 0.11~K and 5~T,
where $\phi^\ast$ is the azimuthal angle in the field-rotational plane.
(b) The same data obtained by converting $\phi^\ast$ to $\phi$ (see text), 
where $\phi$ is the azimuthal angle in the $bc$ plane.
}
\label{phi2}
\end{figure}

Figure \ref{phi2}(a) compares the field-angle $\phi^\ast$ dependences of the $c$-axis and $a$-axis magnetostrictions measured 
under a magnetic field rotated around the $a$ axis and the measurement direction (roughly parallel to the $a$ axis), respectively.
Here, $\phi^\ast$ is the azimuthal angle in the field-rotational plane measured from the direction in which the rotating magnetic field is closest to the $c$ axis.
It was found that the superconducting transition field angle in $\Delta L_a(\phi^\ast)$ does not match with that in $\Delta L_c(\phi^\ast)$.
If we assume $\phi={\rm acos}[\cos\phi^\ast \cos(10.6^\circ)]$ ($\phi=\phi^\ast$) for the $L_a$ ($L_c$) measurements,
an onset field angle of the superconducting transition becomes consistent between $L_c(\phi)$ and $L_a(\phi)$ [see Fig. \ref{phi2}(b)].
Therefore, the measurement direction for $L_a$ seems to be unexpectedly tilted from the $a$ to $c$ axis by 10.6$^\circ$,
possibly due to the irregular shape of the sample;
in this study, minimally polished as-grown samples were used to avoid cracking and distortion of the samples that could easily induce the $\Bc2$ distribution.

\begin{figure}
\includegraphics[width=5.5in]{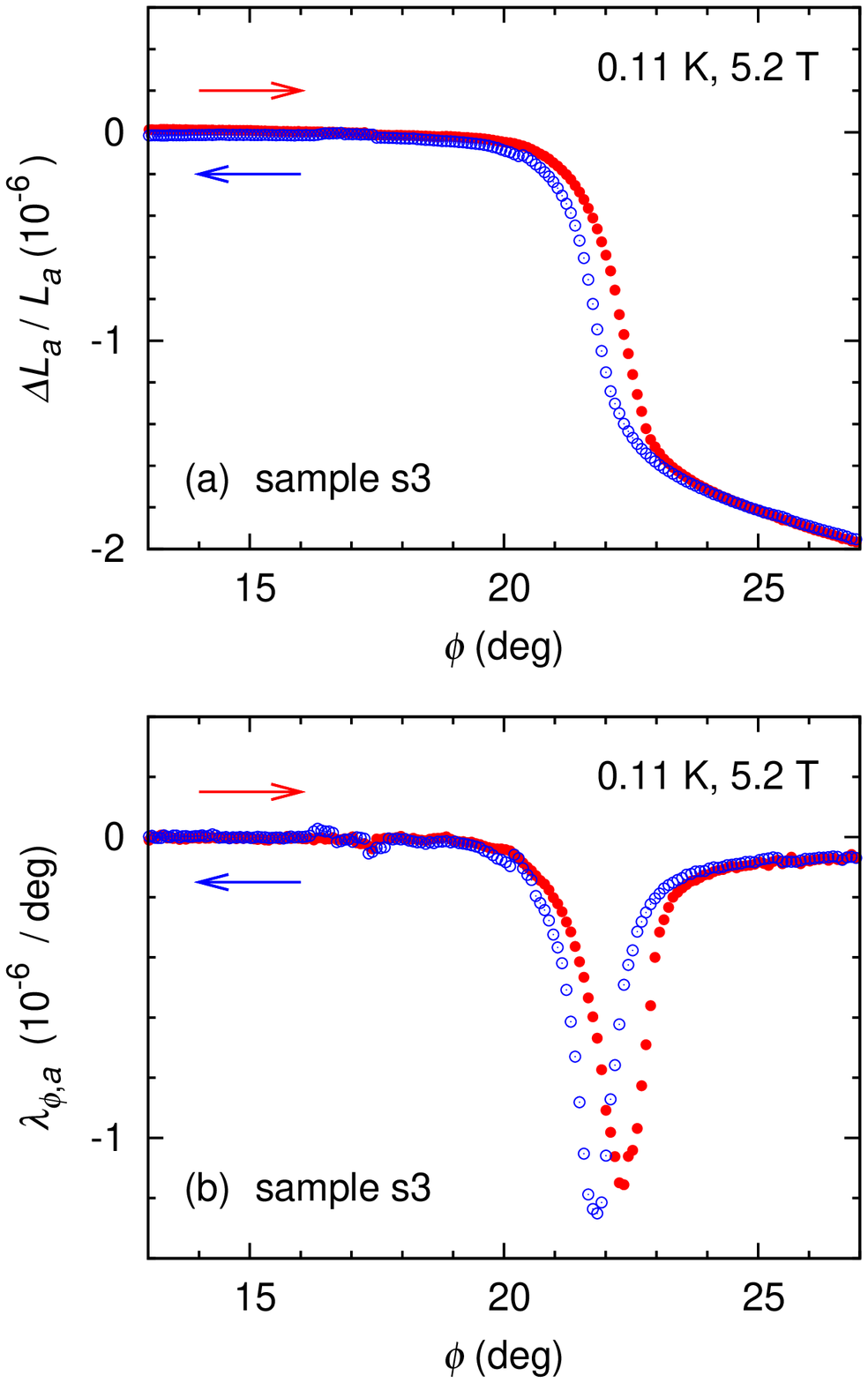}
\caption{
Field-angle $\phi$ dependences of (a) $\Delta L_a/L_a$ and (b) $\lambda_{\phi,a}$ of the sample s3 at 0.11~K and 5.2~T.
}
\label{phia}
\end{figure}

As already explained in the main text, the second anomaly is absent in $L_a(\phi)$. 
The same conclusion can be obtained when the magnetic field is increased up to 5.2~T (Fig. \ref{phia}).
Similar to the large anomaly in $\lambda_{\phi,c}$ (Fig.~\ref{phic}), the anomaly in $\lambda_{\phi,a}$ depends on the direction of the rotating field,
supporting that the small second anomaly observed in $\lambda_{\phi,c}$ is absent in $\lambda_{\phi,a}$.

\begin{figure}
\includegraphics[width=6.5in]{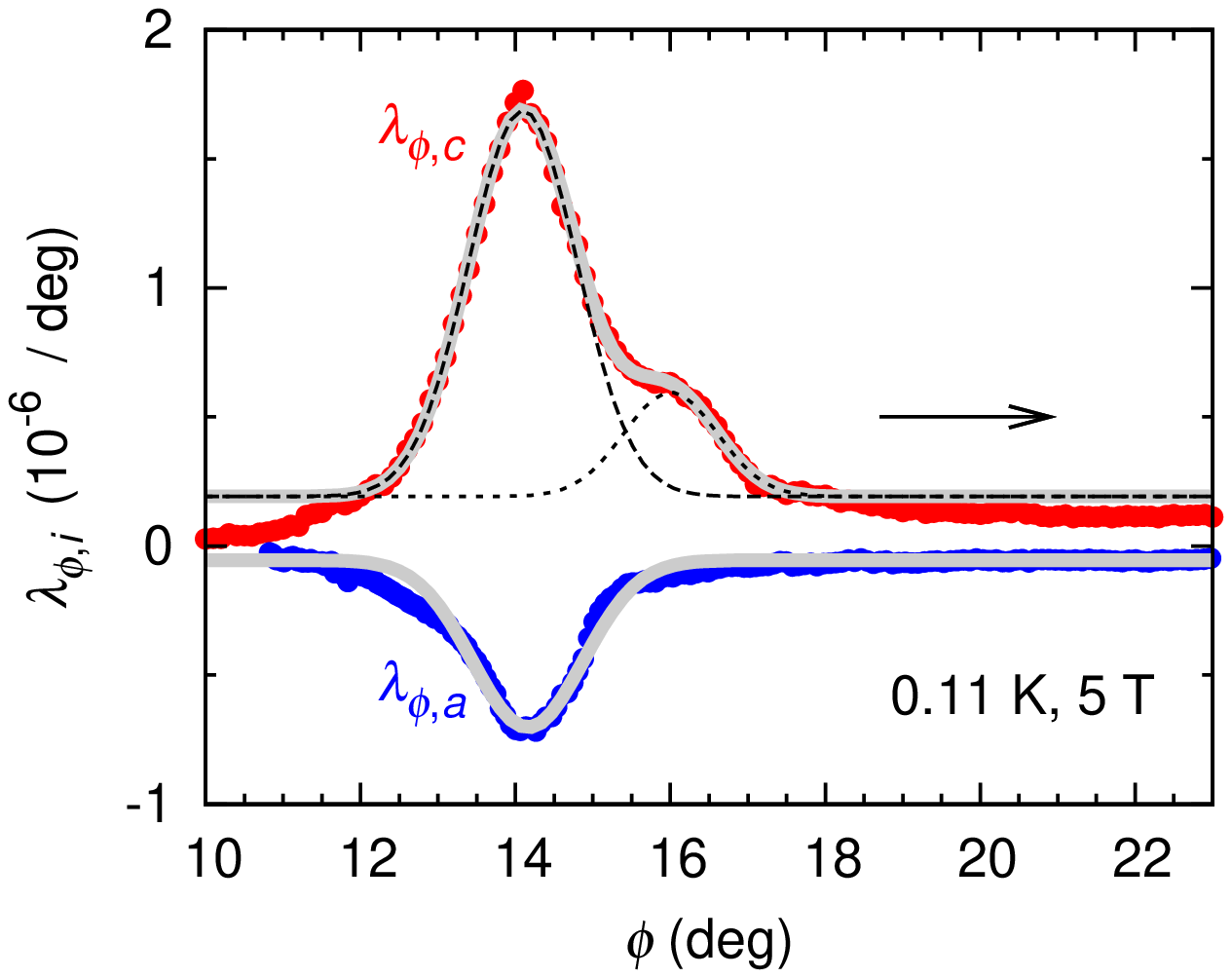}
\caption{
Field-angle $\phi$ dependence of $\lambda_{\phi,i}$ ($i=a$ or $c$) of the sample s3 at 0.11~K and 5~T in the $\phi$-increasing process, 
compared with the Gaussian fits (see text).
}
\label{phifit}
\end{figure}

To characterize the two anomalies at $B_{\rm K}$ and $\Bc2$, we fit to the data of $\lambda_{\phi,c}$ by using a double-Gaussian function 
\begin{equation}
f(\phi)=A_1\exp\biggl[-\frac{(\phi-\phi_1)^2}{2c_1^2}\biggl]+A_2\exp\biggl[-\frac{(\phi-\phi_2)^2}{2c_2^2}\biggl]+f_0
\end{equation}
in the range $12^\circ\le\phi\le18^\circ$, where a double peak is observed.
We also fit to the data of $\lambda_{\phi,a}$ by using the same function with $A_2=0$ in the range $13^\circ\le\phi\le15^\circ$, where a single peak is observed.
The fitting results are represented by solid lines in Fig.~\ref{phifit}. 
Here, for $\lambda_{\phi,c}$, the first- and second-term contributions are represented by dashed and dotted lines.
From these fits, the full width at half maximum (FWHM) can be estimated to be $2\sqrt{2\ln 2}c_j$ ($j=1$ or $2$).
For $\lambda_{\phi,c}$, the FWHM of the $B_{\rm K}$ anomaly is $1.42^\circ\pm0.05^\circ$ whereas that of the $\Bc2$ anomaly is $1.65^\circ\pm0.02^\circ$.
It should be noted that the former value is smaller than the latter one. 
This result eliminates a possibility that 
the $B_{\rm K}$ anomaly is caused by sample inhomogeneities that exist in regions not detected from the $L_a$ measurements 
because a decrease of $\Bc2$ by impurities and/or defects is accompanied by a broadening of the $\Bc2$ distribution as well as suppression of the first-order nature.
For $\lambda_{\phi,a}$, the FWHM of the single anomaly is $1.67^\circ\pm0.03^\circ$, 
which matches well with the FWHM of the $\Bc2$ anomaly in $\lambda_{\phi,c}$, 
further evidencing the absence of the $B_{\rm K}$ anomaly in $\lambda_{\phi,a}$.

\begin{figure}
\includegraphics[width=6.5in]{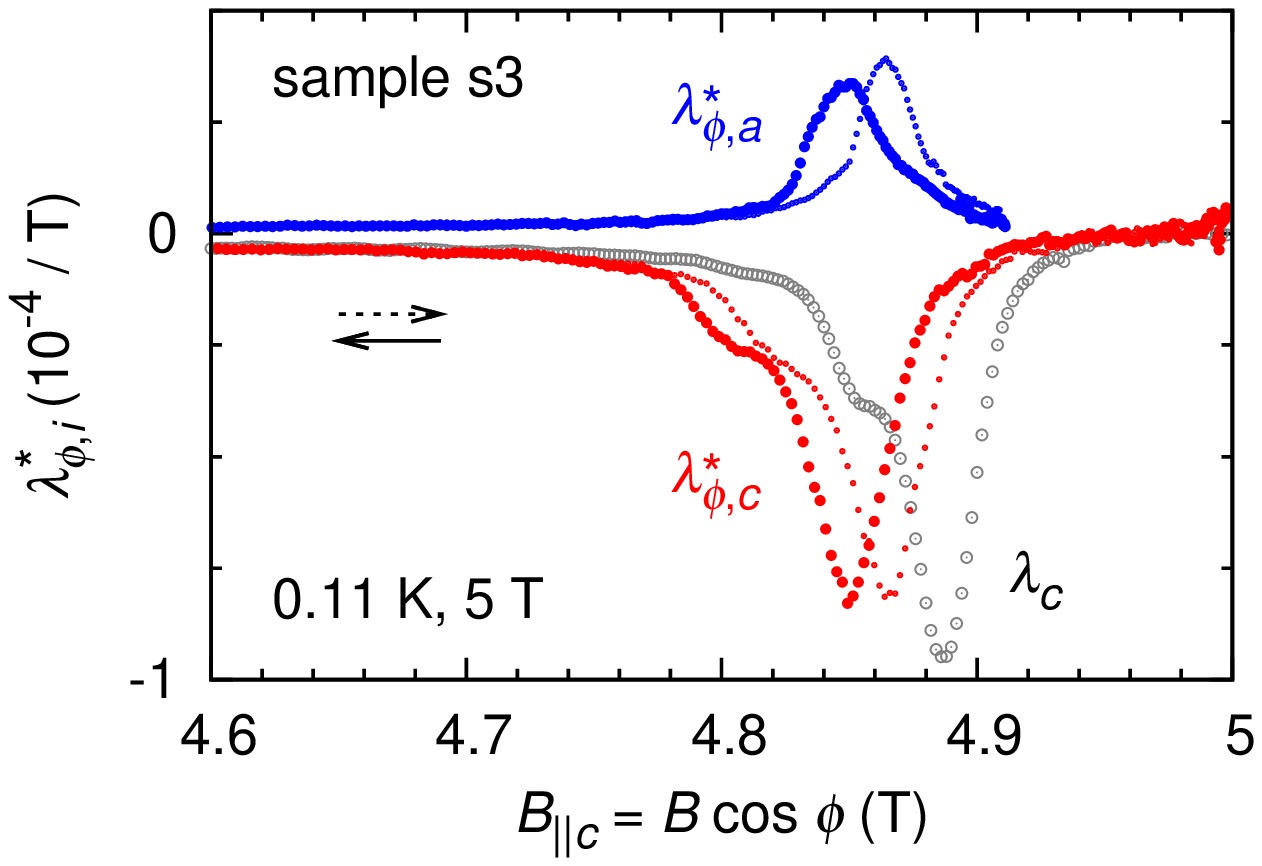}
\caption{
Field-angle-resolved $L_i(\phi)$ data of the sample s3 plotted as a function of the $c$-axis component of $B$, i.e., $B_{\parallel c}=B\cos\phi$.
Large and small closed circles are $\lambda^*_{\phi,i}=[\partial L_i(\phi)/\partial B_{\parallel c}]/L_i$ ($i=a$ or $c$) for $\phi>0$ 
in the $\phi$-increasing and decreasing processes (i.e., $B_{\parallel c}$-decreasing and increasing processes), respectively.
Open circles in grey are $\lambda_c(B)$ for $B \parallel c$ at 0.11~K in the field-increasing process (Fig.~1 in the main text) for comparison.
}
\label{Bc}
\end{figure}

Due to the sample misalignment, 
the magnetic field cannot be applied along the $c$ axis precisely during the $L_a$ measurements.
However, the $c$-axis component of $B$, i.e., $B_{\parallel c}=B\cos\phi$, may be predominant to determine the magnetostriction in this field-angle range 
because of the large anisotropy in $\Bc2$ of CeCoIn$_5$.
Figure \ref{Bc} shows $\lambda^\ast_{\phi,i}=[\partial L_i(\phi)/\partial B_{\parallel c}]/L_i$ ($i=c$ or $a$) as a function of $B_{\parallel c}$ by using the data of Fig.~4(a) of the main text for $\phi>0$.
Here, the data of $\lambda_c(B)$ taken at 0.11~K in the field-increasing process (Fig.~1 of the main text) are also plotted. 
As shown in Fig.~\ref{Bc}, $\lambda^\ast_{\phi,c}(B_{\parallel c})$ essentially reproduces the field dependence of $\lambda_{c}(B)$ at the same temperature, 
except for the smaller $\Bc2$ caused by the in-plane magnetic field component in $\lambda^\ast_{\phi,c}(\phi)$ measurements.
This qualitative agreement between $\lambda^\ast_{\phi,c}(\phi)$ and $\lambda_{c}(B)$ shows that the $B_{\rm K}$ anomaly is also absent in $\lambda_{a}(B)=(\partial L_a/\partial B)/L_a$ for $B \parallel c$.
Thus, the sample misalignment does not affect our key finding that the $B_{\rm K}$ anomaly is not detected from the magnetostriction measurements 
when the measurement direction is perpendicular to the applied magnetic field near $B \parallel c$.

\clearpage

\end{document}